\documentclass[reprint,amsmath,amssymb,aps]{revtex4-2}
\usepackage{float}
\usepackage{appendix}
\usepackage{lipsum}
\usepackage[colorlinks, citecolor=blue,urlcolor=blue,linkcolor=blue,runcolor=filecolor]{hyperref}
\usepackage{url}
\usepackage{amsmath}
\usepackage{subfigure}
\allowdisplaybreaks[4]
\usepackage{graphicx}
\usepackage{epsfig}
\usepackage{dcolumn}
\usepackage{bm}
\usepackage{amsfonts}
\usepackage{amssymb}
\usepackage{array}
\usepackage{graphics}
\usepackage{hyperref}
\usepackage{natbib}
\begin{document}

\title{Constraining the spin-gravity coupling effects to the $10^{-10}$-level with dual-species atom interferometers}

\author{Dongfeng Gao\textsuperscript{1,2,}}
\altaffiliation{Email: dfgao@wipm.ac.cn}
\author{Lin Zhou\textsuperscript{1,2}}
\author{Jin Wang\textsuperscript{1,2,3,}}
\altaffiliation{Email: wangjin@wipm.ac.cn}
\author{Mingsheng Zhan\textsuperscript{1,2,3,}}
\altaffiliation{Email: mszhan@wipm.ac.cn}

\vskip 0.5cm
\affiliation{1 State Key Laboratory of Magnetic Resonance and Atomic and Molecular Physics, Wuhan Institute of Physics and Mathematics, Innovation Academy for Precision Measurement Science and Technology, Chinese Academy of Sciences, Wuhan 430071, China\\
	2 Hefei National Laboratory, Hefei 230088, China\\
	3 Wuhan Institute of Quantum Technology, Wuhan 430206, China}

\date{\today}

\begin{abstract}
	
Spin is one fundamental property of microscopic particles. A lot of theoretical work has postulated the possible coupling between spin and gravitation, which could result in the violation of equivalence principle. In our recent joint mass-and-energy test of the weak equivalence principle with a 10-meter $^{85}$Rb-$^{87}$Rb dual-species atom interferometer, the E${\rm \ddot{o}}$tv${\rm \ddot{o}}$s parameters of four $^{85}$Rb-$^{87}$Rb combinations with specific atomic spin states were measured to the $10^{-10}$-level (\textit{L. Zhou et al., Phys. Rev. A 104, 022822}). Here these experimental results are used to constrain the postulated spin-gravity coupling effects. The bounds on the spin-independent and spin-dependent anomalous passive gravitational mass tensors in L${\rm \ddot{a}}$mmerzahl's model are set to the $10^{-10}$-level, which improves existing bounds by three orders of magnitude. The constraints to the spin-independent electron- and proton-gravity coupling parameters in the gravitational standard-model extension are set to the $10^{-6}\, {\rm GeV}$-level.

\end{abstract}

\maketitle



\section{\label{introduction}Introduction}

The Einstein equivalence principle is the foundation of general relativity, which contains the weak equivalence principle (WEP), the local Lorentz invariance (LLI), and the local position invariance (LPI) \cite{Will2014}. The WEP, also called the universality of free fall, states that the trajectory of a free falling test mass does not depend on its internal structure and composition. In other words, two different bodies will fall with the same acceleration in an external gravitational field. Violations of the WEP are expected in theoretical models on new physics beyond general relativity and the standard model of particle physics \cite{RevModPhys.90.025008}. Most experiments on the WEP test are done with macroscopic test objects in different methods, including torsion balance on laboratory tables \cite{PhysRevLett.121.261101}, lunar laser ranging \cite{Hofmann_2018}, and satellites in space \cite{PhysRevLett.129.121102}, which have achieved $10^{-15}$-level precision.

In recent years, rapid technological progresses in atom interferometry have been made \cite{RevModPhys.81.1051}. Atoms have well-defined and easily manipulated quantum properties, such as atomic spin, the superposition state, and the entanglement state. Therefore, the atom interferometer (AI) turns out to be an ideal tool to test the WEP from various quantum aspects. To perform the test, one has to drop two different atomic species simultaneously, and measure the gravitational acceleration of them. By comparing the difference in the values of the gravitational acceleration, the WEP can be tested. In the past 20 years, both isotopic and nonisotopic atom pairs were used in the WEP test  \cite{PhysRevLett.93.240404,PhysRevA.88.043615,PhysRevLett.112.203002,PhysRevLett.115.013004,Barrett2016,Albers2020,Zhang2020,PhysRevLett.113.023005,PhysRevLett.117.023001,Rosi2017,PhysRevLett.125.191101,zhou2021}. These WEP tests involve different quantum properties, such as bosonic and fermionic nature \cite{PhysRevLett.113.023005}, the superposition state \cite{Rosi2017}, and atom pairs with specific atomic spin combinations \cite{zhou2021}.

On the other hand, it is well known that general relativity was originally formulated as a theory to describe the gravitational interactions between macroscopic masses. These massive bodies are made of microscopic particles, which are labeled by both mass and spin. Then, if one attempts to extend the concepts of general relativity to microscopic particles, the role of spins has to be taken into consideration. Extensive theoretical efforts have been made to investigate the nature of interactions among spins and masses \cite{Ni_2010}. Novel spin-dependent interactions can appear either through constructing models which attempt to combine general relativity with quantum theories (for example, see Refs. \cite{PhysRev.136.B1542,PhysRevLett.36.393,RevModPhys.48.393, PhysRevD.18.2739, Lammerzahl1998,PhysRevLett.86.192,Mashhoon_2000,DonatoBini_2004,PhysRevD.70.103515,PhysRevD.83.016013,PhysRevD.103.024059, PhysRevD.104.044054}), or by introducing some postulated mediation particles leading to various spin-dependent potentials (for example, see Refs. \cite{Naik1981,PhysRevD.30.130, Dobrescu_2006, PhysRevD.80.105021}). In the meantime, many experimental tests of the possible spin-gravity coupling effects have been performed, which are based on WEP tests with AIs \cite{PhysRevLett.113.023005, PhysRevLett.117.023001}, polarized macroscopic massive objects \cite{Ni_2010, PhysRevD.78.092006}, stored-ion spectroscopy \cite{PhysRevLett.67.1735}, and atomic magnetometers \cite{PhysRevLett.68.135, PhysRevD.96.075004, PhysRevLett.120.161801, PhysRevLett.130.201401}. Especially, both AI experiments aiming for the spin-gravity coupling effects, one performed with the fermionic-bosonic $^{87}$Sr-$^{88}$Sr pair \cite{PhysRevLett.113.023005} and the other performed with two different hyperfine states of the $^{87}$Rb atom \cite{PhysRevLett.117.023001}, have achieved the $10^{-7}$-level WEP test.

In this paper, we work on L${\rm \ddot{a}}$mmerzahl's model \cite{Lammerzahl1998} and the gravitational standard-model extension (SME) \cite{PhysRevD.83.016013,PhysRevD.103.024059, PhysRevD.104.044054}, where certain non-relativistic spin-gravity couplings were discussed. Based on our recent atomic WEP test experiment where the E${\rm \ddot{o}}$tv${\rm \ddot{o}}$s parameters of four $^{85}$Rb-$^{87}$Rb combinations were measured to the $10^{-10}$-level \cite{zhou2021}, we study the possibility of using these experimental results to constrain spin-gravity coupling parameters. For L${\rm \ddot{a}}$mmerzahl's model, the bounds on the atomic spin-gravity coupling parameters are set to the $10^{-10}$-level. For the gravitational SME, the bounds on the spin-independent electron- and proton-gravity coupling parameters are set to the $10^{-6}\, {\rm GeV}$-level. The paper is organized as follows. In Sec. \ref{the shift phase}, the WEP test with dual-species AIs is briefly discussed. In Sec. \ref{constrain the coupling parameters}, we discuss how to constrain the spin-gravity coupling parameters in  L${\rm \ddot{a}}$mmerzahl's model, using the newest atomic WEP tests. In Sec. \ref{the SME},  we discuss how to constrain the spin-gravity couplings in the gravitational SME. Finally, discussion and conclusion are made in Sec. \ref{conlusion and discussion}. The detailed calculations for L${\rm\ddot{a}}$mmerzahl's model and the gravitational SME are discussed in the appendix \ref{AppendixA} and \ref{AppendixB}.

\section{The WEP test with dual-species atom interferometers}
\label{the shift phase}

\begin{figure}[h]
	\centering
	\includegraphics[width=9cm,height=5cm]{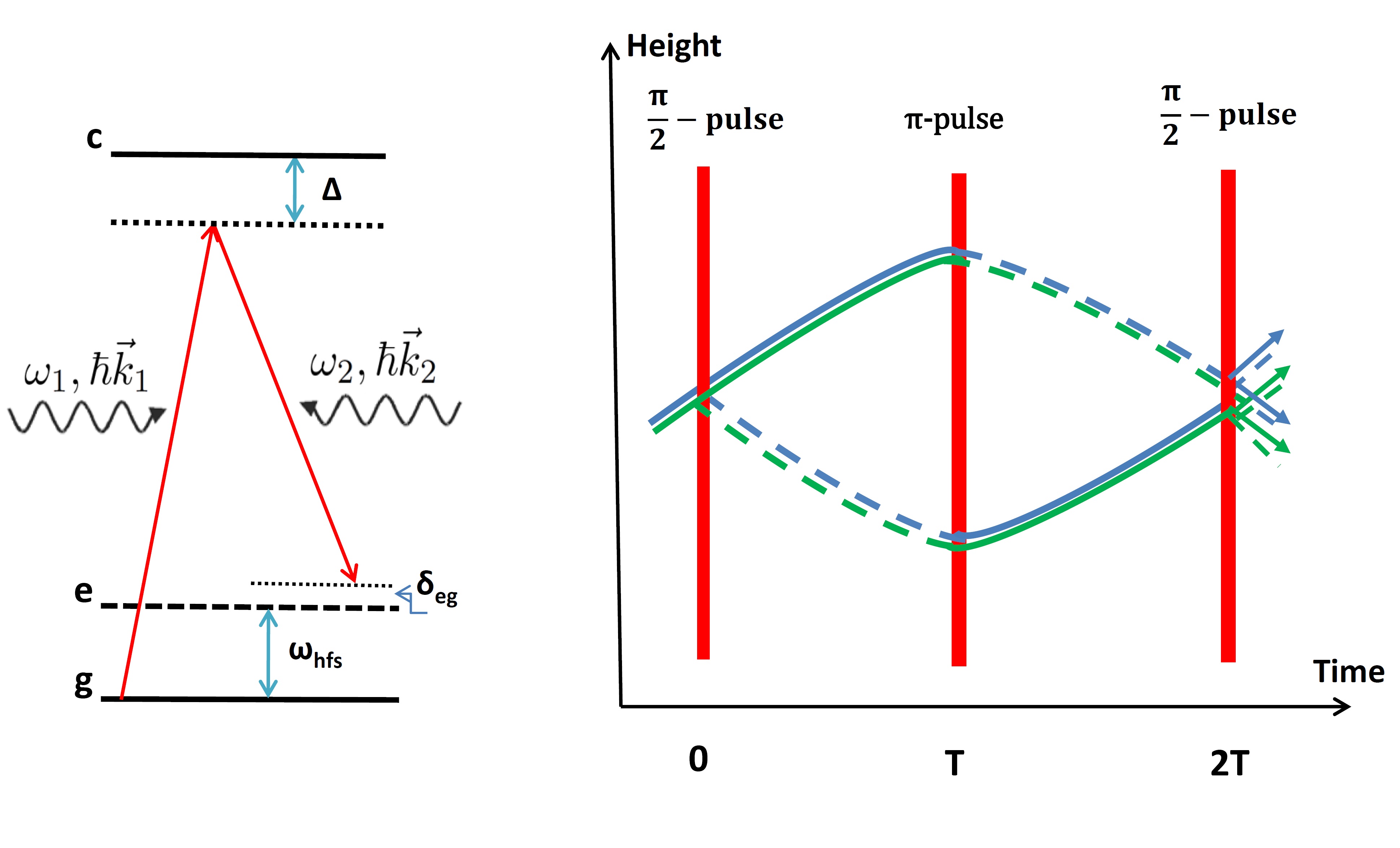}
	\caption{Schematic diagram for a typical $\frac{\pi}{2}$-$\pi$-$\frac{\pi}{2}$ Raman atom interferometer. The left part shows the stimulated Raman transition between two atomic hyperfine ground states $|g\rangle$ and $|e\rangle$. The right part shows the sequence of laser pulses and the paths of atoms, where blue and green colors denote the two different atomic species.}
	\label{figure1}
\end{figure}

The theory of AIs has been reviewed in many papers, such as Ref. \cite{Kasevich1992}. A typical $\frac{\pi}{2}$-$\pi$-$\frac{\pi}{2}$ Raman AI is depicted in Fig. \ref{figure1}. The cold atom beam, prepared in the $|g\rangle$ state, is first coherently split into a superposition of states $|g\rangle$ and $|e\rangle$ (with a momentum difference of $\hbar \vec{k}$=$\hbar(\vec{k}_1$-$\vec{k}_2)$) at time $t=0$. After a drift time T, Raman $\pi$-pulses are applied to transit the states $|g\rangle$ to $|e\rangle$, and $|e\rangle$ to $|g\rangle$, respectively. After another drift time T, the two wave packets overlap and interfere. Finally, the phase shift for each atomic species can be measured by detecting the number of atoms in either $|g\rangle$ or $|e\rangle$ states.

If the AI is located in a gravitational field $\vec{g}$,  then it is straightforward to calculate the phase shift
\begin{equation}
	\Phi=-\vec{k} \cdot \vec{g} T^2.
	\label{phaseshift}
\end{equation}
Once $\Phi$ is measured with the AI, given that $\vec{k}$ and $T$ are measured with other instruments, the gravitational acceleration can thus be determined.

To test the WEP, the accelerations of two different atomic species in the same gravitational field have to be measured simultaneously. Then, the technique of dual-species atom interferometry is developed, where two atomic species can be either isotopic atoms (such as the $ ^{87}{\rm Sr}$-$ ^{88}{\rm Sr}$ pair used in Ref. \cite{PhysRevLett.113.023005}, and the $ ^{85}{\rm Rb}$-$ ^{87}{\rm Rb}$ pair used in Refs. \cite{zhou2021,PhysRevLett.93.240404,PhysRevA.88.043615,PhysRevLett.125.191101}) or nonisotopic atoms (such as the $ ^{87}{\rm Rb}$ and $ ^{39}{\rm K}$ atoms used in Ref. \cite{PhysRevLett.112.203002,Barrett2016}). By the simultaneous realization of Raman interference for both atomic species (as depicted in Fig. \ref{figure1}), their gravitational accelerations, denoted by $g_1$ and $g_2$, can be measured simultaneously. Thus, the E${\rm \ddot{o}}$tv${\rm \ddot{o}}$s parameter, defined through
\begin{equation}
	\eta=\frac{g_1-g_2}{(g_1+g_2)/2}\,
\end{equation}
can be calculated and analyzed. The best result for $\eta$ with dual-species AIs was given in Ref. \cite{PhysRevLett.125.191101}, which is $1.6\pm 1.8({\rm stat})\pm 3.4({\rm syst})\times 10^{-12}$.

However, for the purpose of this paper, we will focus on the measurement in Ref. \cite{zhou2021}, where the authors proposed a four-wave double-diffraction Raman transition scheme, so that they could measure the E${\rm \ddot{o}}$tv${\rm \ddot{o}}$s parameters of four $^{85}$Rb-$^{87}$Rb combinations with specific atomic spin states. For $^{85}$Rb atoms, 5S$_{1/2}$ $|$F=2, $F_z$=0$\rangle$ and 5S$_{1/2}$ $|$F=3, $F_z$=0$\rangle$ were used. To avoid confusion, they were abbreviated as $^{85}$Rb$|$2$\rangle$ and $^{85}$Rb$|$3$\rangle$, respectively.  For $^{87}$Rb atoms, 5S$_{1/2}$ $|$F=1, $F_z$=0$\rangle$ (abbreviated as $^{87}$Rb$|$1$\rangle$) and 5S$_{1/2}$ $|$F=2, $F_z$=0$\rangle$ (abbreviated as $^{87}$Rb$|$2$\rangle$) were used. Note that $F$ denotes the atomic spin number, and $F_z$ denotes the z-component of the atomic spin. Then, four different $^{85}$Rb-$^{87}$Rb combinations were prepared and the $\eta$ for each combination was measured. The measured values of $\eta$ were the following,
\begin{eqnarray}
	\nonumber	{\rm ^{85}Rb|2\rangle-^{87}Rb|1\rangle}:  &  &  \,\,\,\,\,\eta_1=(1.5\pm 3.2)\times 10^{-10},\\
	\nonumber	{\rm ^{85}Rb|2\rangle-^{87}Rb|2\rangle}:  &  &  \,\,\,\,\,\eta_2=(-0.6\pm 3.7)\times 10^{-10},\\
	\nonumber	{\rm ^{85}Rb|3\rangle-^{87}Rb|1\rangle}:  &  & \,\,\,\,\, \eta_3=(-2.5\pm 4.1)\times 10^{-10},\\
\nonumber	{\rm ^{85}Rb|3\rangle-^{87}Rb|2\rangle}:  &  & \,\,\,\,\, \eta_4=(-2.7\pm 3.6)\times 10^{-10}.
\end{eqnarray}

\section{\label{constrain the coupling parameters}Constraints on the spin-gravity coupling parameters in L${\rm \ddot{a}}$mmerzahl's model}


The model \cite{Lammerzahl1998} and the corresponding phase shift in AI experiments are discussed in Appendix \ref{AppendixA}. For our case, the phase shift is
\begin{equation}
	\Phi = - \left(1 +\alpha\right) k\, g\, T^2
	\label{phaseshiftc}
\end{equation}
with
\begin{eqnarray}
	\alpha
	& = & {{\delta m_{\hbox{\scriptsize P}}^{\,\,zz}}\over m} - {{\delta m_{\hbox{\scriptsize I}}^{\,zz}}\over m}- 2 \left({{\delta \bar m_{\hbox{\scriptsize I}k}}^{zz}\over m} + C_k\right) F^k\, ,	
\end{eqnarray}
where $F^k$ denotes the k-component of the atomic spin. $\delta m_{\hbox{\scriptsize I}}^{zz}$ and $\delta \bar m_{\hbox{\scriptsize I} k}^{zz}$ stand for the spin-independent and spin-dependent anomalous inertial mass tensors. $\delta m_{\hbox{\scriptsize P}}^{\,\,zz}$ gives the spin-independent anomalous passive gravitational mass tensors. $C_i$ is the spin-dependent anomalous passive gravitational mass vector.

Although all the $^{85}$Rb and $^{87}$Rb atomic beams were prepared in the $F^z$=0 states, they carry an inherent quantum fluctuation in their spin components
\begin{equation*}
	\Delta F_i = \sqrt{\langle F|F_i^2|F\rangle-\langle F|F_i|F\rangle^2}\, .
\end{equation*}
For this reason, the measured coefficient $\alpha$ in our AI experiments is actually
\begin{eqnarray}
	{\bar \alpha}
	& = & {{\delta m_{\hbox{\scriptsize P}}^{\,\,zz}}\over m} - {{\delta m_{\hbox{\scriptsize I}}^{\,zz}}\over m}- 2 \left({{\delta \bar m_{\hbox{\scriptsize I}k}}^{zz}\over m} + C_k\right) \Delta F^k\, .	
\end{eqnarray}
It is easy to calculate that $\Delta F_x=\Delta F_y=1$ for F=1, $\Delta F_x=\Delta F_y=\sqrt{3}$ for F=2 and $\Delta F_x=\Delta F_y=\sqrt{6}$ for F=3.

According to L${\rm \ddot{a}}$mmerzahl's model, the spin-gravity coupling effects contribute to the inertial mass and the passive gravitational mass of atoms. Thus, the inherent quantum fluctuation in spin components induces quantum fluctuations in the inertial mass and the passive gravitational mass. In the end, this results in the violation of the equivalence principle. For the $^{85}$Rb-$^{87}$Rb dual-species AIs, the WEP violation can be written as
\begin{eqnarray*}
	\eta & &= {\bar \alpha}_{85}-{\bar \alpha}_{87}\\
	& &= \left( {{\delta m_{\hbox{\scriptsize P}}^{\,\,zz}}\over m} - {{\delta m_{\hbox{\scriptsize I}}^{\,zz}}\over m}\right)_{85}-\left( {{\delta m_{\hbox{\scriptsize P}}^{\,\,zz}}\over m} - {{\delta m_{\hbox{\scriptsize I}}^{\,zz}}\over m}\right)_{87} \\
	& &- 2\left. \left({{\delta \bar m_{\hbox{\scriptsize I}k}}^{zz}\over m} + C_k\right) \Delta F^k\right|_{85}+2\left. \left({{\delta \bar m_{\hbox{\scriptsize I}k}}^{zz}\over m} + C_k\right) \Delta F^k\right|_{87}
\end{eqnarray*}
Note that a similar argument was also used in Ref. \cite{Adunas2001}. Inserting the four measured values of $\eta$, we eventually find constraints for the following four combinations of model parameters:
\begin{eqnarray}
\nonumber &&\left( {{\delta m_{\hbox{\scriptsize P}}^{\,\,zz}}\over m} - {{\delta m_{\hbox{\scriptsize I}}^{\,zz}}\over m}\right)_{85}-\left( {{\delta m_{\hbox{\scriptsize P}}^{\,\,zz}}\over m} - {{\delta m_{\hbox{\scriptsize I}}^{\,zz}}\over m}\right)_{87}\\
\nonumber && \,\,\,\,\,\,\,\,\,\,\,\,\,\,\,\,\,\,\,\,\,\,\,\,\,\,\,\,\,\,\,\,\,\,\,\,\,\,\,\,\,\,\,\,\,\,\,\,\,\,\,\,\,\,\,\,\,\,\,\, =  (1.4\pm 1.9)\times 10^{-9}\\	
\nonumber &&\left({{\delta \bar m_{\hbox{\scriptsize I}x}}^{zz}+{\delta \bar m_{\hbox{\scriptsize I}y}}^{zz}}\over m \right)_{85}= (0.7\pm 2.5)\times 10^{-8}\\
\nonumber &&\left({{\delta \bar m_{\hbox{\scriptsize I}x}}^{zz}+{\delta \bar m_{\hbox{\scriptsize I}y}}^{zz}}\over m \right)_{87}= (0.6\pm 2.5)\times 10^{-8}\\	
&&\,\,\,\,\,\,\,\,\,\,\,\,\,\,\,\,\,\,\,\,\,\,\,\,\,\,\,\,\,\,\,\,\,\,\, C_x+C_y =  (-0.6\pm 2.5)\times 10^{-8}\, .
\label{result1}
\end{eqnarray}

Compared to the estimations given in Ref. \cite{Lammerzahl1998}, our results improve the constraints by about four orders of magnitude. If we look into the constraints (\ref{result1}) carefully, it seems that we might lose one or two orders of magnitude. The reason is that the LLI-violation (${\delta \bar m_{\hbox{\scriptsize I}i}}^{jk}$ and $\delta m_{\hbox{\scriptsize I}}^{\,jk}$) and LPI-violation (${\delta m_{\hbox{\scriptsize P}}^{\,\,jk}}/ m$ and $C_k$) effects are mixed up and added up, which results in a loss of sensitivity.

\begin{figure}[h]
	\centering
	\includegraphics[width=7.5cm,height=5cm]{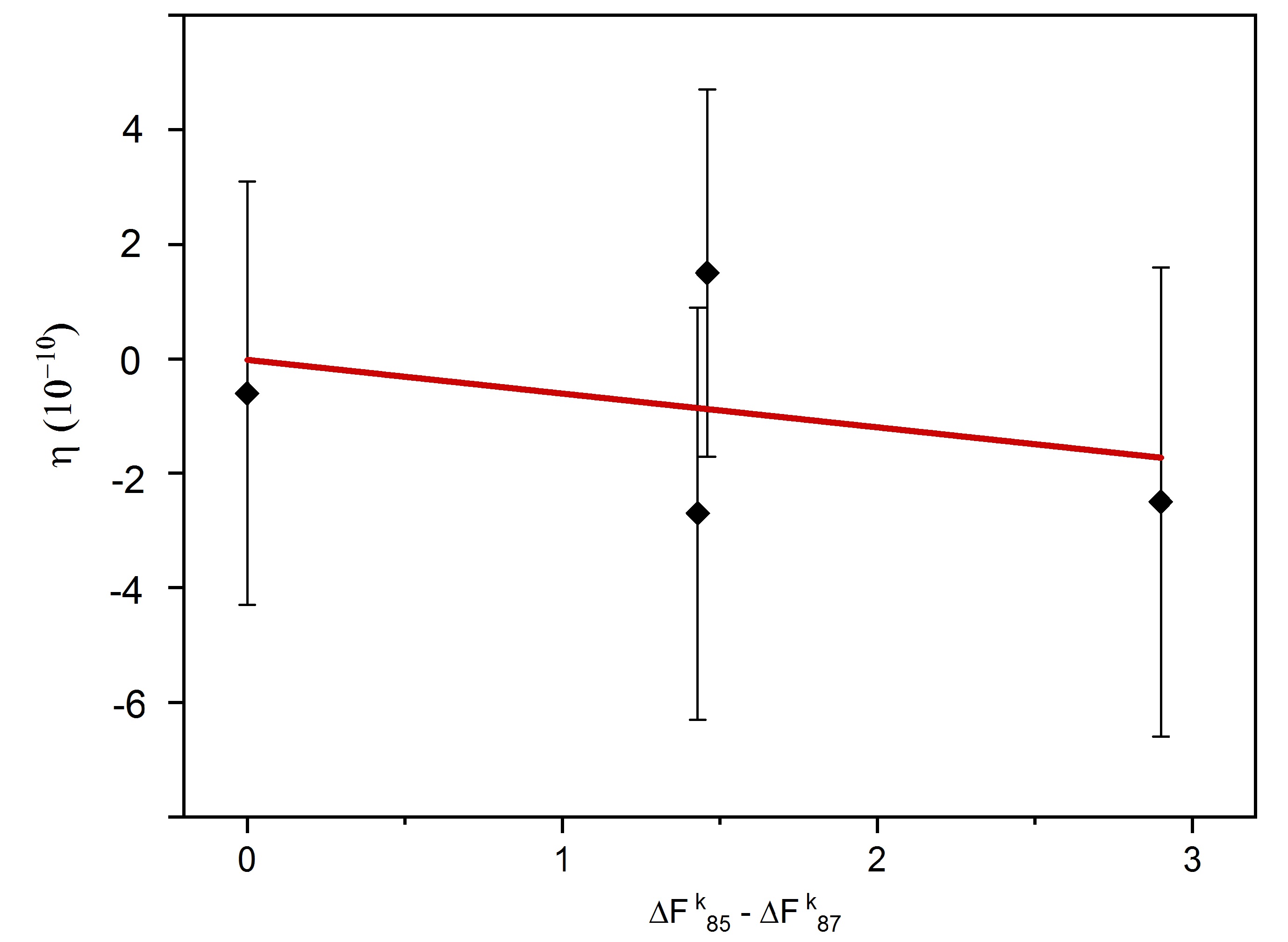}
	\caption{The intercept value of the fitted straight line, $({\delta m_{\hbox{\scriptsize P}}^{\,\,zz}}/ m)_{85}$-$({\delta m_{\hbox{\scriptsize P}}^{\,\,zz}}/ m)_{87}$=$(0.0\pm 2.1)\times 10^{-10}$ and the slope, $C_x$+$C_y$ = $(-0.6\pm 1.3)\times 10^{-10}$. }
	\label{figure2}
\end{figure}

 Actually, we could get better constraints on LPI-violation parameters. As discussed in Ref. \cite{Lammerzahl1998}, Hughes-Drever-type experiments \cite{PhysRevLett.54.2387,PhysRevLett.57.3125,PhysRevLett.63.1541} are far more sensitive on LLI-violation effects than atom interferometry, and tighter bounds, $|{\delta \bar m_{\hbox{\scriptsize I}i}}^{jk}/m|$ and $|{\delta m_{\hbox{\scriptsize I}}^{\,jk}}/m|\sim 10^{-30}$, have already been set. Thus, let us neglect the LLI-violation effects in atom interferometry. The E${\rm \ddot{o}}$tv${\rm \ddot{o}}$s parameter $\eta$ is then simplified into
\begin{eqnarray*}
	\eta
	& &\simeq \left( {{\delta m_{\hbox{\scriptsize P}}^{\,\,zz}}\over m}\right)_{85}-\left( {{\delta m_{\hbox{\scriptsize P}}^{\,\,zz}}\over m}\right)_{87}- 2C_k \left(\left. \Delta F^k\right|_{85}- \left. \Delta F^k\right|_{87}\right)
\end{eqnarray*}
We can use this equation to fit the four measured values of $\eta$ into a straight line, shown in Fig. \ref{figure2}. One can easily read out that
\begin{eqnarray}
	\nonumber &&\left( {{\delta m_{\hbox{\scriptsize P}}^{\,\,zz}}\over m}\right)_{85}-\left( {{\delta m_{\hbox{\scriptsize P}}^{\,\,zz}}\over m}\right)_{87} =  (0.0\pm 2.1)\times 10^{-10}\\		
	&&\,\,\,\,\,\,\,\,\,\,\,\,\,\,\,\,\,\,\,\, C_x+C_y =  (-0.6\pm 1.3)\times 10^{-10}
\label{result2}
\end{eqnarray}
Compared with Eq. (\ref{result1}), the constraint on $C_x+C_y$ is improved by two orders of magnitude. It clearly shows that the constraints on the LPI-violation effects are improved once they are singled out from other effects.

Let us further focus on the spin-independent LPI-violation parameter, ${\delta m_{\hbox{\scriptsize P}}^{\,\,zz}}/ m$. As explained in Refs. \cite{Will2014, 10.1007/3-540-40988-2_10},  LPI-violation effects can be best tested by the gravitational null redshift experiments done with two atomic clocks (A and B). The violation parameter $\Delta \alpha^{RS}$, used in null redshift experiments, is related to the spin-independent LPI-violation parameter by
\begin{equation*}
	\Delta \alpha^{RS} = ({\delta m_{\hbox{\scriptsize P}}^{\,\,zz}}/ m)_{A}-({\delta m_{\hbox{\scriptsize P}}^{\,\,zz}}/ m)_{B} \, .
\end{equation*}
Currently, the best bounds come from comparing the relative frequency ratios of the $^{87}$Rb and $^{133}$Cs atomic fountains \cite{PhysRevLett.109.080801} giving $\Delta \alpha^{RS}=(0.1\pm 1.0)\times 10^{-6}$, and two different isotopes of Dysprosium \cite{PhysRevLett.111.060801} giving $\Delta \alpha^{RS}=(-5.5\pm 5.2)\times 10^{-7}$. Our result (\ref{result2}) improves the bounds by three orders of magnitude.

\section{\label{the SME} Constraints on the spin-gravity couplings in the gravitational SME}

As discussed in Appendix \ref{AppendixB}, the gravitational SME attributes the couplings between atoms and gravity to the constituent electrons, protons, and neutrons in atoms \cite{PhysRevD.103.024059, PhysRevD.104.044054}. Consider an atom of mass $m_{atom}$ consisting of $N_e$ electrons, $N_p$ protons, and $N_n$ neutrons, located in a gravitational field $\vec{g}$. The spin-gravity induced correction to the gravitational acceleration of the free falling atom is found to be
\begin{eqnarray}
	\delta \vec{g} 
	& &=
	\sum_{w,N_w} \Big[
	(k^{{\rm NR}}_{\phi})_w + (k^{{\rm NR}}_{\phi pp})^{jk}_w p^j_w p^k_w \nonumber\\
	& &
	+ (k^{{\rm NR}}_{\sigma\phi})_w^j \sigma^j_w +(k^{{\rm NR}}_{\sigma\phi pp})_w^{jkl} \sigma^j_w p^k_w p^l_w \Big] \frac{\vec{g}}{m_{atom}},
	\label{operator}
\end{eqnarray}
where the index $w$ stands for e, n, and p. The coefficients $(k^{{\rm NR}}_{\phi})_w$ and $(k^{{\rm NR}}_{\phi pp})^{jk}_w$ denote the spin-independent couplings to gravity via the gravitational potential $\phi$. The coefficients $(k^{{\rm NR}}_{\sigma\phi})_w^j$ and $(k^{{\rm NR}}_{\sigma\phi pp})_w^{jkl}$ denote the spin-dependent couplings to gravity via $\phi$. 

For both $^{85}$Rb and $^{87}$Rb atoms, there is a single valence electron in the $5^2S_{1/2}$ level,
with the total electronic angular momentum $J^e=1/2$. In the Schmidt model \cite{Blatt1979}, the nuclear properties for both $^{85}$Rb and $^{87}$Rb atoms are assigned to a single valence proton. The nucleus of $^{85}$Rb atoms has nuclear spin $I=5/2$ with orbital angular momenta $L^p=3$ \cite{PhysRevC.10.2117}. The nucleus of $^{87}$Rb atoms has nuclear spin $I=3/2$ with orbital angular momenta $L^p=1$ \cite{PhysRevC.12.1516}. Through a bit of calculation, the expectation values of Eq. (\ref{operator}) for the four states are found to be
\begin{widetext}
	\begin{eqnarray}
		\left. {\delta g \over g}\right|_{{\rm ^{85}Rb}|3\rangle}= \left. {\delta g \over g}\right|_{{\rm ^{85}Rb}|2\rangle}&=&{1\over m_{85}}\left[ (k^{{\rm NR}}_{\phi})_e+(k^{{\rm NR}}_{\phi})_p+\left( {9\over 35}\left((k^{{\rm NR}}_{\phi pp})_p^{xx}+(k^{{\rm NR}}_{\phi pp})_p^{yy}\right)+{17\over 35}(k^{{\rm NR}}_{\phi pp})_p^{zz}\right) \left\langle \vec{p}^{\,2} \right\rangle_p \right] \\
		\label{sme85}
		\left.{\delta g \over g}\right|_{{\rm ^{87}Rb}|2\rangle} =\left. {\delta g \over g}\right|_{{\rm ^{87}Rb}|1\rangle}&=&{1\over m_{87}}\left[ (k^{{\rm NR}}_{\phi})_e+(k^{{\rm NR}}_{\phi})_p+\left( {4\over 15}\left((k^{{\rm NR}}_{\phi pp})_p^{xx}+(k^{{\rm NR}}_{\phi pp})_p^{yy}\right)+{7\over 15}(k^{{\rm NR}}_{\phi pp})_p^{zz}\right) \left\langle \vec{p}^{\,2} \right\rangle_p \right]
		\label{sme87}
	\end{eqnarray}	
\end{widetext}
where the subscripts $e$ and $p$ stand for the valence electron and proton, respectively. For typical atoms, the mean momentum squared is of order $\langle\vec{p}^{\, 2}\rangle_p \simeq 10^{-2}$ GeV$^2$. Since we are dealing with $F_z$=0 cases, corrections from the spin-dependent couplings are zero. It turns out that, for the four $^{85}$Rb-$^{87}$Rb combinations, the calculated E${\rm \ddot{o}}$tv${\rm \ddot{o}}$s parameter is 
\begin{widetext}
\begin{eqnarray*}
\eta&=&({1\over m_{85}}-{1\over m_{87}})\left((k^{{\rm NR}}_{\phi})_e+(k^{{\rm NR}}_{\phi})_p\right)+\left[({9\over 35 m_{85}}-{4\over 15 m_{87}})
\left((k^{{\rm NR}}_{\phi pp})_p^{xx}+(k^{{\rm NR}}_{\phi pp})_p^{yy}\right)+({17\over 35 m_{85}}-{7\over 15 m_{87}})(k^{{\rm NR}}_{\phi pp})_p^{zz}\right]\left\langle \vec{p}^{\,2} \right\rangle_p
\end{eqnarray*}
\end{widetext}
Inserting the averaged value ($(-1.1\pm 3.6)\times 10^{-10}$) of the four measured values of $\eta$ in Ref. \cite{zhou2021}, one can get the following bounds  
\begin{eqnarray}
(k^{{\rm NR}}_{\phi})_e+(k^{{\rm NR}}_{\phi})_p&=&   (-0.4\pm 1.4)\times 10^{-6}\, {\rm GeV} \nonumber\\
\label{sme111}
	(k^{{\rm NR}}_{\phi pp})_p^{xx}+(k^{{\rm NR}}_{\phi pp})_p^{yy}&=& (2.4\pm 8.0)\times 10^{-4}\, {\rm GeV^{-1}} \\	
	(k^{{\rm NR}}_{\phi pp})_p^{zz}&=& (-3.0\pm 9.7)\times 10^{-5}\, {\rm GeV^{-1}}\nonumber
\end{eqnarray}
Note that we assume only one combination of model parameters to be nonzero at a time. 

As discussed in Ref. \cite{PhysRevD.104.044054}, WEP tests with $^{87}$Sr atoms of different spins \cite{PhysRevLett.113.023005}, and with $^{87}$Rb atoms in different hyperfine states \cite{PhysRevLett.117.023001} can be used to constrain the spin-dependent couplings. As a complement to their constraints, our results (\ref{sme111}) set constraints to the spin-independent sector of the gravitational SME. The classic COW experiment \cite{PhysRevLett.34.1472} can be used to constrain the spin-independent coupling for neutrons, which is $(k^{{\rm NR}}_{\phi})_n< 1\times 10^{-2}\, {\rm GeV}$. Our results give the $10^{-6}\, {\rm GeV}$-level bound for protons. From the viewpoint of nucleons, our results are complementary to the experiments with neutrons.

\section{\label{conlusion and discussion} Conclusion and Discussion}

L${\rm \ddot{a}}$mmerzahl's model \cite{Lammerzahl1998} and the gravitational SME \cite{PhysRevD.83.016013,PhysRevD.103.024059, PhysRevD.104.044054} provide theoretical frameworks to study both spin-independent and spin-dependent LLI- and LPI-violation effects, which are clearly related to the occurrence of the WEP-violation. Dual-species atom interferometry based on matter waves in specific atomic spin states is an ideal tool to test possible spin related WEP-violations. The bounds on L${\rm \ddot{a}}$mmerzahl's model parameters are set to the $10^{-10}$-level, which is three orders of magnitude better than the one set by null redshift experiments \cite{PhysRevLett.109.080801, PhysRevLett.111.060801}. The constraints to the spin-independent electron- and proton-gravity coupling parameters in the gravitational SME are set to the $10^{-6}\, {\rm GeV}$-level, which is complementary to the experiments with neutrons.

Processes towards higher precision in dual-species atom interferometry are being made. As mentioned before, the $10^{-12}$-level WEP test had already been achieved for the $^{85}$Rb$|$3$\rangle$-$^{87}$Rb$|$2$\rangle$ combination \cite{PhysRevLett.125.191101}, which would give a $10^{-12}$-level bound on $({\delta m_{\hbox{\scriptsize P}}^{\,\,zz}}/ m)_{85}$-$({\delta m_{\hbox{\scriptsize P}}^{\,\,zz}}/ m)_{87}$ if one could ignore all other LPI-violation parameters. Our research group is working on pushing the WEP test to the $10^{-11}$-level and higher for all four $^{85}$Rb-$^{87}$Rb combinations \cite{zhou2022}. Moreover, some space-based proposals, such as STE-QUEST \cite{PhysRevD.109.064010}, plan to push  the WEP test to the $10^{-17}$-level with atom interferometry. Thus, one could expect to set better bounds on spin-gravity coupling effects in the near future.

\begin{acknowledgements}
This work was supported by the Technological Innovation 2030 "Quantum Communication and Quantum Computer" Major Project (Grants No. 2021ZD0300603 and No. 2021ZD0300604), the National Natural Science Foundation of China (Grants No. 12241410 and No. 12174403), the Project for Young Scientists in Basic Research of CAS (Grant No. YSBR-055),  the Hubei Provincial Science and Technology Major Project (Grant No. ZDZX2022000001), and the Natural Science Foundation of Hubei Province under (Grant No. 2022CFA096).

\end{acknowledgements}

\appendix
\section{The spin-gravity couplings in L${\rm \ddot{a}}$mmerzahl's model \label{AppendixA}}

\subsection{L${\rm \ddot{a}}$mmerzahl's model}

In this appendix, we briefly discuss the spin-gravity model proposed in Ref. \cite{Lammerzahl1998}. For a neutral massive fermionic particle, the model starts with the generalized Dirac equation (GDE)  
\begin{equation}
	i \partial_t\varphi = - i c (\widetilde\alpha^i \partial_i
	+ i \Gamma) \varphi + m c^2 \widetilde\beta \varphi  \label{GDE}
\end{equation}
with $i, j = 1, 2, 3$. $\varphi$ is a complex four-component spinor field. The $4\times 4$-matrices ($\widetilde\alpha^i$, $\widetilde\beta^i$ and $\Gamma$) obey $(\widetilde\alpha^i)^+ = \widetilde\alpha^i$, $\widetilde\beta^+ = \widetilde\beta$, and $\Gamma^+ =\Gamma + i \partial_i \widetilde\alpha^i$. Note that they are not assumed to fulfill a Clifford algebra.

In the non-relativistic limit, two components of the generalized Dirac equation become small. By eliminating these small components, the GDE will reduce to the two-component Pauli equation. 

To do the non-relativistic limit, one just follows the general procedure as usual. Use the matrix,  $\beta = \begin{pmatrix} 1 & 0 \\ 0 & -1 \end{pmatrix}$, to distinguish the upper and lower (i.e., the large and small) components. Define the even and odd operators $\cal E$ and $\cal O$ through $\beta {\cal E} = {\cal E} \beta$ and $\beta {\cal O} = - {\cal O} \beta$. 

One first formally splits $\widetilde\alpha^i$, $\widetilde\beta$ and $\Gamma$ into even and odd parts: $\widetilde\alpha^i = \widetilde\alpha^i_e + \widetilde\alpha^i_o$,  $\widetilde\beta = \widetilde\beta_e + \widetilde\beta_o$ and $\Gamma = \Gamma_e + \Gamma_o$. Then, the GDE can be written into 
\begin{eqnarray}
	i \partial_t\varphi & = & - i c (\widetilde\alpha^i_e + \widetilde\alpha^i_o) \partial_i \varphi + c \Gamma \varphi + m c^2 \widetilde\beta \varphi \nonumber\\
	& = & \beta m c^2 + {\cal E} + {\cal O}
\end{eqnarray}
with
\begin{eqnarray}
	{\cal O} & = & - i c \widetilde\alpha^i_o \partial_i + c \Gamma_o + m c^2 \widetilde\beta_o \\ 
	{\cal E} & = & - i c \widetilde\alpha^i_e \partial_i + c \Gamma_e + m c^2 \widetilde\beta_e - m c^2 \beta
\end{eqnarray}

Now, performing a Foldy-Wouthuysen transformation \cite{PhysRev.78.29} with $\varphi^\prime = U \varphi$, $U = e^{i S}$, and $S = - {i\over{2 m}} \beta {\cal O}$, we get  
\begin{eqnarray}
	H^\prime\varphi^\prime & & =  \beta \left(m c^2 + {{{\cal O}^2}\over{2 m c^2}} - {{{\cal O}^4}\over{8 m^3 c^6}}\right)\varphi^\prime + {\cal E}\varphi^\prime \nonumber\\
	& & - {1\over{8 m^2 c^4}} \left[{\cal O}, [{\cal O}, {\cal E}]\right]\varphi^\prime - {i\over{8 m^2 c^4}} [{\cal O}, \dot{\cal O}]\varphi^\prime 
	\label{GPE1}
\end{eqnarray}

For the case where the fermionic particle is located in a gravitational field, the author made the following ansatz,
\begin{eqnarray*}
	\widetilde\beta(x) & = & {\buildrel 0 \over {\widetilde\beta}} + {\buildrel 1 \over {\widetilde\beta}}{1\over{c^2}} \left(U(x) + {{\delta m^\prime_{\hbox{\scriptsize P} kl}}\over m} U^{kl}(x)\right) \label{widetildebeta} \\ 
	\widetilde\alpha^i(x) & = & {\buildrel 0 \over {\widetilde\alpha}}{}^i + {\buildrel 1 \over {\widetilde\alpha}}{}^i {1\over{c^2}} \left(U(x) + {{\delta m^\prime_{\hbox{\scriptsize P} kl}}\over m} U^{kl}(x)\right) \\ 
	\Gamma(x) & = & {\buildrel 0 \over \Gamma} + {\buildrel 1 \over \Gamma} {1\over{c^2}} \left(U(x) + {{\delta m^\prime_{\hbox{\scriptsize P} kl}}\over m}U^{kl}(x)\right) \\
	& &+ {\buildrel 1 \over \Gamma}{}^i {1\over{c^2}} \left(\partial_i U(x) + {{\delta m^\prime_{\hbox{\scriptsize P} kl}}\over m} \partial_i U^{kl}(x)\right) 
\end{eqnarray*}
where ${\buildrel 0 \over {\widetilde\beta}}$, ${\buildrel 1 \over {\widetilde\beta}}$, ${\buildrel 0 \over {\widetilde\alpha}}{}^i$, ${\buildrel 1 \over {\widetilde\alpha}}{}^i$, ${\buildrel 0 \over \Gamma}$, ${\buildrel 1 \over \Gamma}$, and ${\buildrel 1 \over \Gamma}{}^i$ are constant matrices. $U(x)$ and $U^{ij}(x)$ are the Newtonian potential and the gravitational tensor for the gravitational field sourced by a mass density $\rho (x)$.

Furthermore, since Eq. (\ref{GPE1}) involves products of $\widetilde\alpha^i_{o,e}$, $\widetilde\beta_{o,e}$ and $\Gamma_{o,e}$ matrices, we have to use the projection operator $P = {1\over 2}(1 + \beta)$ to pick up the large components. By doing the projection, the author introduced the following terms 
\begin{eqnarray*}
	P {\buildrel 0 \over {\widetilde\alpha}}{}^{(i}_o {\buildrel 0 \over {\widetilde\alpha}}{}^{j)}_o & = & \delta^{ij} + {{\delta m_{\hbox{\scriptsize I}}^{ij}}\over m} + {{\delta \bar m_{\hbox{\scriptsize I} k}^{ij}}\over m} \sigma^k \label{anomalinert}\\   
	P {\buildrel 0 \over {\widetilde\alpha}}{}^i_e & = & A^i + A^i_j \sigma^j \\  
	P \{{\buildrel 0 \over \Gamma}_o, {\buildrel 0 \over {\widetilde\alpha}}{}^i_o\}  & = & 2 (a^i + a^i_j \sigma^j) \\ 
	P {\buildrel 0 \over {\Gamma_e}} & = & T + T_j \sigma^j \\
	P {\buildrel 1 \over {\widetilde\beta}}_e & = & 1 + d + C_j \sigma^j \\  
	P({\buildrel 0 \over {\widetilde\beta}}_e - \beta) & = & B + B_j \sigma^j \\  
	P{\buildrel 0 \over {\widetilde\alpha}}{}^{[i}_o {\buildrel 0 \over {\widetilde\alpha}}{}^{j]}_o & = & K^{ij} + (\epsilon^{ij}_{\phantom{ij}k} + K^{ij}_k) \sigma^k 
\end{eqnarray*}
where $\sigma^i$ are the usual Pauli matrices. One can see that each matrix product is projected into a scalar part and a spin part. 

Doing a further transformation on Eq. (\ref{GPE1}) by means of $\varphi^\prime = e^{- i m c \delta_{ij} (A^i + {1\over m} a^i) x^j} \widehat \varphi^\prime$, we finally arrive at the generalized Pauli equation
\begin{eqnarray}
	H^\prime \widehat\varphi^\prime 
	& = & - {1\over{2 m}} \left(\delta^{ij} + {{\delta m_{\hbox{\scriptsize I}}^{ij}}\over m} + {{\delta \bar m_{\hbox{\scriptsize I} k}^{ij}}\over m} \sigma^k\right) \partial_i \partial_j \widehat\varphi^\prime  \nonumber\\
	& &	-i \left({1\over m} a^i_j + c A^i_j\right) \sigma^j \partial_i \widehat\varphi^\prime  + \Biggl[(m c^2 B_i + c T_i) \sigma^i \nonumber\\
	& & + (1 + C_i \sigma^i) m U(x) + \delta m_{\hbox{\scriptsize P} ij} U^{ij}(x)\Biggr] \widehat\varphi^\prime \label{GPE}
\end{eqnarray}
where $\delta m_{\hbox{\scriptsize P} ij}=\delta m^\prime_{\hbox{\scriptsize P} ij}+d$. Now, the physical meanings of the previously introduced parameters become clear. $\delta m_{\hbox{\scriptsize I}}^{ij}$ and $\delta \bar m_{\hbox{\scriptsize I} k}^{ij}$ stand for the spin-independent and spin-dependent anomalous inertial mass tensors. $a^i_j$ and $A^i_j$ give the spin-momentum couplings. $m c^2 B_i$ can be regarded as a spin-dependent rest mass, and $T_i$ can be interpreted as the space-like part of an axial torsion vector. $\delta m_{\hbox{\scriptsize P} ij}$ gives the spin-independent anomalous passive gravitational mass tensor. $C_i$ is the spin-dependent anomalous passive gravitational mass vector. To summarize, $\delta m_{\hbox{\scriptsize I}}^{ij}$, $\delta\bar m_{\hbox{\scriptsize I} k}^{ij}$, $a^i_j$, $A^i_j$, and $B_i$ stand for the LLI violation, while $C_i$ and $\delta m_{\hbox{\scriptsize P}ij}$ result in the LPI violation. If all these parameters vanish, then we go back to the usual Schr${\rm \ddot{o}}$dinger equation.

Finally, it is straightforward to write down the corresponding classical Hamiltonian 
\begin{eqnarray*}
	H^\prime & = & {p_i p_j \over{2 m}} \left(\delta^{ij} + {{\delta m_{\hbox{\scriptsize I}}^{ij}}\over m} + 2 {{\delta\bar m_{\hbox{\scriptsize I} k}^{ij}}S^k\over m} \right) - 2 \left({a^i_j \over m} + c A^i_j\right) S^j p_i \nonumber\\
	& & + 2 (m c^2 B_i + c T_i) S^i + (1 + 2 C_i S^i) m U + \delta m_{\hbox{\scriptsize P} ij} U^{ij}
\end{eqnarray*}
where $p_i$ and $S^i$ are the momentum and spin of the particle, respectively. Then, it is straightforward to obtain the velocity, force and acceleration for the particle
\begin{eqnarray}
	v^i & = & {1\over m} \left(\delta^{ij} + {{\delta m_{\hbox{\scriptsize I}}^{ij}}\over m} + 2 {{\delta\bar m_{\hbox{\scriptsize I} k}^{ij}}\over m} S^k\right) p_j - 2 \left({1\over m} a^i_j + c A^i_j\right) S^j \nonumber\\
	f_i & = & - (1 + 2 C_j S^j) m \partial_i U - \delta m_{\hbox{\scriptsize P} kl} \partial_i U^{kl} \nonumber\\ 
	a^i & = & - \left(\delta^{ij} + {{\delta m_{\hbox{\scriptsize I}}^{ij}}\over m} + 2 \left({{\delta\bar m_{\hbox{\scriptsize I} k}^{ij}}\over m} + \delta^{ij} C_k\right) S^k\right) \partial_j U \nonumber\\
	& &  - \delta^{ij} {{\delta m_{\hbox{\scriptsize P} kl}}\over m} \partial_j U^{kl} \label{accel}
\end{eqnarray}
where the dynamics of the spin vector was neglected. Although the above results are derived for the spin-1/2 particles, they also hold for higher spin particles by carrying out a similar consideration.

\subsection{The phase shift in AI experiments}

To calculate the AI phase shift in the spin-gravity model \cite{Lammerzahl1998}, we have to first calculate the acceleration (\ref{accel}) for the Earth. 
Following the Ref. \cite{Will1993}, we introduce several useful gravitational potentials
\begin{eqnarray}
	U(\vec{r}, t) & = &G \int \frac{\rho(\vec{r},t)}{|\vec{r}-\vec{r}'|} d^3 r'
\end{eqnarray}
\begin{eqnarray}
	\chi(\vec{r}, t) & = &G \int \rho(\vec{r}',t)|\vec{r}-\vec{r}'| d^3 r'
\end{eqnarray}
\begin{eqnarray}
	U_{ij}(\vec{r}, t) & = &G \int \frac{\rho(\vec{r}',t)(\vec{r}-\vec{r}')_i({\vec{r}}-{\vec{r}'})_j}{|{\vec{r}}-{\vec{r}'}|^3} d^3 r'
\end{eqnarray}
where G is the universal gravitational constant, and $\rho(\vec{r},t)$ is the mass density of the Earth.
They satisfy
\begin{eqnarray*}
	\chi_{, ij} &=& \delta_{ij}U-U_{ij}, \,\,\,\,\,\,\,\,\,\delta^{ij}U_{ij}=U, \,\,\,\, {\rm with}\,\,\,\,\, \partial_i \equiv \frac{\partial}{\partial r^i} 
\end{eqnarray*}
It is easy to find that 
\begin{eqnarray*}
	\chi(\vec{r}, t) & = & G M_E \left(r + \frac{R_E^2}{5r}\right)
\end{eqnarray*}
\begin{eqnarray*}
	U(\vec{r}, t) & = & \frac{G M_E}{r}, \,\,\,\,\,\,\,\,\, {\rm with}\,\,\,\,\, r \equiv (r_1^2 +r_2^2+r_3^2)^{1/2}
\end{eqnarray*}
where $M_E$ and $R_E$ are the mass and the radius of the Earth, respectively.  
Notice that
\begin{eqnarray*}
	\partial_i \, r^n= n\, r^{n-2} \, r_i\, ,
\end{eqnarray*}
\begin{eqnarray*}
	\partial_i \, r&=& \frac{r_i}{r} , \,\,\,\,\,\,\,\,\, \partial_i \partial_j\, r=\frac{\delta_{ij}}{r}- \frac{r_i r_j}{r^3}\\
	\partial_i \, r^{-1}&=& \frac{-r_i}{r^3} , \,\,\,\,\,\,\,\,\, \partial_i \partial_j\, r^{-1}=-\frac{\delta_{ij}}{r^3}+ \frac{3 r_i r_j}{r^5}\\
\end{eqnarray*}
One can easily find out 
\begin{eqnarray*}
	\nonumber	U_{ij}(\vec{r}, t) & = &  \delta_{ij}U - \chi_{, ij}\\ 
	&=& \frac{G M_E}{r^3} \left(\frac{R_E^2}{5}\delta_{ij}+r_i r_j-\frac{3 R_E^2}{5}\frac{ r_i r_j}{r^2}\right)
\end{eqnarray*}
It is straightforward to calculate
\begin{eqnarray*}
	\partial_k U_{ij}	&=& -\frac{3G M_E}{5 r^5}\left(5 r_i r_j r_k+R_E^2(\delta_{ij} r_k+\delta_{ik} r_j+\delta_{jk} r_i)\right) \\
	& & +\frac{G M_E }{r^3}(\delta_{ik} r_j+\delta_{jk} r_i)+\frac{3 G M_E R_E^2}{r^7}r_i r_j r_k
\end{eqnarray*}

Thus, the acceleration (\ref{accel}) is calculated to be
\begin{widetext}
	\begin{eqnarray*}
		\nonumber  a^l&=&- \left(\delta^{lj} + {{\delta m_{\hbox{\scriptsize I}}^{lj}}\over m} + 2 \left({{\delta\bar m_{\hbox{\scriptsize I} k}^{lj}}\over m} + \delta^{lj} C_k\right) S^k\right) \partial_j U - \delta^{kl}{{\delta m_{\hbox{\scriptsize P}}^{\,ij} }\over m}\partial_k U_{ij}\\
		\nonumber	&=&\frac{G M_E}{r^3}\left(r^l + {{\delta m_{\hbox{\scriptsize I}}^{lj}}\over m} r_j+ 2 \left({{\delta\bar m_{\hbox{\scriptsize I} k}^{lj}}\over m} + \delta^{lj} C_k\right) S^k r_j\right)\\
		\nonumber
		& & +\frac{G M_E}{r^3}\left\lbrace \frac{3}{r^2}\frac{\delta m_{\hbox{\scriptsize P}}^{\,ij}}{m} r_i r_j r^l+\frac{3 R_E^2}{5 r^2}\left(\frac{\delta m_{\hbox{\scriptsize P}}^{\,ii}}{m} r_l+\frac{\delta m_{\hbox{\scriptsize P}}^{\,lj}}{m} r_j+\frac{\delta m_{\hbox{\scriptsize P}}^{\,il}}{m} r_i \right) \right. 
		\left.  -\frac{\delta m_{\hbox{\scriptsize P}}^{\,lj}}{m} r_j-\frac{\delta m_{\hbox{\scriptsize P}}^{\,il}}{m} r_i - \frac{3 R_E^2}{r^4}\frac{\delta m_{\hbox{\scriptsize P}}^{\,ij}}{m} r_i r_j r^l\right\rbrace 
	\end{eqnarray*}
\end{widetext}

Suppose $\vec{r}_0$ is the vector from the center of the Earth to the beam splitter. $|\vec{r}_0|=R_E + h$, where $h=\frac{1}{2} g T^2$ is the height of the atom interferometer. Since h is typically of several meters, then $h/R_E \sim 10^{-6}$. Also assume that $\delta m_{\hbox{\scriptsize P}}^{\,ij}$ is symmetric, i.e., $\delta m_{\hbox{\scriptsize P}}^{\,ij}=\delta m_{\hbox{\scriptsize P}}^{\,ji}$. Then, the corresponding phase shift is 
\begin{widetext}
	\begin{eqnarray}
		\Phi & = & T^2 \frac{G M_E}{r_0^3} k_l \left\lbrace \frac{3}{r_0^2}\frac{\delta m_{\hbox{\scriptsize P}}^{\,ij}}{m} {r_0}_i {r_0}_j {r_0}^l+\frac{3 R_E^2}{5 r_0^2}\left(\frac{\delta m_{\hbox{\scriptsize P}}^{\,ii}}{m} {r_0}_l+\frac{\delta m_{\hbox{\scriptsize P}}^{\,lj}}{m} {r_0}_j+\frac{\delta m_{\hbox{\scriptsize P}}^{\,il}}{m} {r_0}_i \right) \right.  \nonumber\\
		& & \left.  -\frac{\delta m_{\hbox{\scriptsize P}}^{\,lj}}{m} {r_0}_j-\frac{\delta m_{\hbox{\scriptsize P}}^{\,il}}{m} {r_0}_i - \frac{3 R_E^2}{r_0^4}\frac{\delta m_{\hbox{\scriptsize P}}^{\,ij}}{m} {r_0}_i {r_0}_j {r_0}^l\right\rbrace +T^2 \frac{G M_E}{r_0^3} k_l \left(r_0^l + {{\delta m_{\hbox{\scriptsize I}}^{lj}}(S)\over m} r_{0j}\right) \nonumber\\
		& \simeq & T^2 \frac{G M_E}{r_0^3} k_l \left\lbrace  \frac{3}{r_0^2}\frac{\delta m_{\hbox{\scriptsize P}}^{\,ij}}{m} {r_0}_i {r_0}_j {r_0}^l + \frac{3}{5}(1-2 h/R_E )\left(\frac{\delta m_{\hbox{\scriptsize P}}^{\,ii}}{m} {r_0}_l+2 \frac{\delta m_{\hbox{\scriptsize P}}^{\,lj}}{m} {r_0}_j \right) \right. \nonumber\\
		& & \left. -2 \frac{\delta m_{\hbox{\scriptsize P}}^{\,lj}}{m} {r_0}_j - \frac{3}{r_0^2}(1-2 h/R_E )\frac{\delta m_{\hbox{\scriptsize P}}^{\,ij}}{m} {r_0}_i {r_0}_j {r_0}^l\right\rbrace +T^2 \frac{G M_E}{r_0^3} k_l \left(r_0^l + {{\delta m_{\hbox{\scriptsize I}}^{lj}}(S)\over m} r_{0j}\right)\nonumber\\
		&=& T^2 {{G M_E}\over{r_0^3}}\left\lbrace
		\frac{6 h}{R_E}{{\delta m_{\hbox{\scriptsize P} ij} }\over m} {{r_0^i r_0^j}\over{r_0^2}}
		k_l r^l_0 - {{4}\over 5}{{\delta m_{\hbox{\scriptsize P} ij}}\over m} r_0^i k^j  +
		{3\over 5} {{\delta m_{\hbox{\scriptsize P} ii}}\over m} k_l r^l_0\right. \nonumber\\
		& & \left. - \frac{6 h}{5 R_E} \left({{\delta m_{\hbox{\scriptsize P} ii}}\over m} k_l r^l_0+2{{\delta m_{\hbox{\scriptsize P} ij}}\over m} r_0^i k^j \right) \right\rbrace +T^2 \frac{G M_E}{r_0^3} k_l \left(r_0^l + {{\delta m_{\hbox{\scriptsize I}}^{lj}}(S)\over m} r_{0j}\right)\nonumber\\
		& \simeq & 	T^2 {{G M_E}\over{r_0^3}}\left( - {{4}\over 5}{{\delta m_{\hbox{\scriptsize P} ij}}\over m} r_0^i k^j  + {3\over 5} {{\delta m_{\hbox{\scriptsize P} ii}}\over m} k_l r^l_0 \right)+T^2 \frac{G M_E}{r_0^3} k_l \left(r_0^l + {{\delta m_{\hbox{\scriptsize I}}^{lj}}(S)\over m} r_{0j}\right)\, ,
		\label{phaseshifta}
	\end{eqnarray}
\end{widetext}
where $\delta m_{\hbox{\scriptsize I}}^{ij}(S) := {\delta m_{\hbox{\scriptsize I}}^{ij}}+ 2 \left({\delta\bar m_{\hbox{\scriptsize I} k}^{ij}} + \delta^{ij} m C_k\right) S^k$, and all terms involving $h/R_E$ have been ignored in the last line. Our result (\ref{phaseshifta}) is slightly different from Eq. (32) in Ref. \cite{Lammerzahl1998}, which is due to some mistake made in Ref. \cite{Lammerzahl1998}.

Furthermore, we align $\vec{r}_0 \sim \vec e_z$ as in usual AI experiments and denote the angle between $\vec{r}_0$ and $\vec{k}$ by $\theta$. Then, to the lowest order in $\delta m_{\hbox{\scriptsize P} ij}$, we get
\begin{equation}
	\Phi = - \left(1 +\alpha\right) k\, T^2 g \cos\left(\theta + b\right)
	\label{phaseshiftb}
\end{equation}
with
\begin{eqnarray}
	\alpha & = & {1\over 5}\left(-{{\delta m_{\hbox{\scriptsize P}}^{\,\,zz}}\over m} + 6 {{\delta m_{\hbox{\scriptsize P}}^{\,\,xx}}\over m}\right) - {{\delta m_{\hbox{\scriptsize I}}^{\,zz}(S)}\over m}\nonumber \\
	& = & {{\delta m_{\hbox{\scriptsize P}}^{\,\,zz}}\over m} - {{\delta m_{\hbox{\scriptsize I}}^{\,zz}}\over m}- 2 \left({{\delta \bar m_{\hbox{\scriptsize I}k}}^{zz}\over m} + C_k\right) S^k	     
\end{eqnarray}	
\begin{eqnarray}			
	b & = & {4\over 5} {{\delta m_{\hbox{\scriptsize P} zx}}\over m} + {{\delta m_{\hbox{\scriptsize I} zx}(S)}\over m} \label{alphab}
\end{eqnarray}
where $\delta m_{\hbox{\scriptsize P}}^{\,\,xx} = \delta m_{\hbox{\scriptsize P}}^{\,\,yy}= \delta m_{\hbox{\scriptsize P}}^{\,\,zz}$ has been used, assuming the spherical symmetry of the mass. It is clearly that $\alpha$ and $b$ involve the diagonal and off-diagonal components of $\delta m_{\hbox{\scriptsize P} ij}$ and $\delta m_{\hbox{\scriptsize I} ij}(S)$ tensors, respectively.
Note that $\alpha$ and $b$ are constant for one species of atoms.

\section{The spin-gravity couplings in the standard-model extension (SME) \label{AppendixB}} 

The spin-gravity coupling effects can also be addressed by the gravitational sector of the (SME \cite{PhysRevD.83.016013,PhysRevD.103.024059, PhysRevD.104.044054}, which was claimed to be model independent. 

\subsection{The gravitational sector in the SME}

Here, we use the notations in Refs. \cite{PhysRevD.103.024059, PhysRevD.104.044054}. The gravitational SME focuses on issues related to violations of local Lorentz and diffeomorphism invariance, and investigates what possible terms are allowed. The whole discussion is comprehensive and complicated. For our purpose, we focus on the couplings that obey the LLI but violate the LPI. The non-relativistic results are quoted briefly as follows. 

First, let us write down the non-relativistic Hamiltonian $H$ of the gravitational SME
\begin{equation}
	H=H_0 + H_{\phi} + H_{\sigma\phi} + H_{g} + H_{\sigma g} + \cdots,
	\label{hamiltonian}
\end{equation}
where $H_0$ is the Hamiltonian in the Minkowski background. The subscript $\sigma$ stands for the spin-dependent terms. The subscript $\phi$ stands for the gravitational potential $\phi$, and the gravitational acceleration is defined by $\vec g \equiv -\vec \nabla \phi$. For applications to laboratory experiments, one can typically write down
\begin{equation*}
	H_0 = \frac{\vec{p}^{\,2}}{2m}-m\vec{g}\cdot\vec{z}-\frac{3}{4m}(\vec{p}^{\,2}\vec{g}\cdot\vec{z}
	+\vec{g}\cdot\vec{z}~\vec{p}^{\,2})+ \frac{3}{4m} (\vec{\sigma}\times\vec{p})\cdot\vec{g}.
\end{equation*}
$H_{\phi}$ denotes the spin-independent couplings via the gravitational potential $\phi$, which is
\begin{eqnarray}
	H_{\phi}&=&(k^{{\rm NR}}_{\phi}) \vec{g}\cdot\vec{z}
	+(k^{{\rm NR}}_{\phi p})^j \frac{1}{2}(p^j \vec{g}\cdot\vec{z}
	+\vec{g}\cdot\vec{z}\ p^j) \nonumber\\
	&&
	+(k^{{\rm NR}}_{\phi pp})^{jk} \frac{1}{2}(p^j p^k \vec{g}\cdot\vec{z}
	+\vec{g}\cdot\vec{z}\ p^j p^k),
	\label{hphi}
\end{eqnarray}
The spin-dependent couplings via the gravitational potential $\phi$ are given by
\begin{eqnarray}
	H_{\sigma\phi}&=&(k^{{\rm NR}}_{\sigma\phi})^j \sigma^j \vec{g}\cdot\vec{z}
	+(k^{{\rm NR}}_{\sigma\phi p})^{jk} \frac{1}{2} \sigma^j (p^k \vec{g}\cdot\vec{z}
	+\vec{g}\cdot\vec{z}\ p^k)
	\nonumber\\
	&&
	+(k^{{\rm NR}}_{\sigma\phi pp})^{jkl} \frac{1}{2} \sigma^j (p^k p^l \vec{g}\cdot\vec{z}
	+\vec{g}\cdot\vec{z}\ p^k p^l).
	\label{hsigmaphi}
\end{eqnarray}
$H_{g}$ denotes the spin-independent couplings via the gravitational acceleration $\vec g$, which
can be written into the form
\begin{equation}
	H_{g}=(k^{{\rm NR}}_{g})^j g^j+(k^{{\rm NR}}_{g p})^{jk} p^j g^k 
	+(k^{{\rm NR}}_{gpp})^{jkl}p^j p^k g^l.
	\label{hg}
\end{equation}
The spin-independent couplings via the gravitational acceleration $\vec g$ are found to be 
\begin{eqnarray}
	H_{\sigma g}&=&(k^{{\rm NR}}_{\sigma g})^{jk} \sigma^j g^k
	+(k^{{\rm NR}}_{\sigma g p})^{jkl} \sigma^j p^k g^l 
	\nonumber\\
	&&
	+(k^{{\rm NR}}_{\sigma g p })^{jklm} \sigma^j p^k p^l g^m,
	\label{hsigmag}
\end{eqnarray}

Next, consider an atom of mass $m_{atom}$ consisting of $N_e$ electrons, $N_p$ protons, and $N_n$ neutrons, located in a gravitational field $\vec{g}$. The spin-gravity induced correction to the gravitational acceleration of the free falling atom is found to be
\begin{eqnarray}
	\delta \vec{g} 
	& &=
	\sum_{w,N_w} \Big[
	(k^{{\rm NR}}_{\phi})_w + (k^{{\rm NR}}_{\phi pp})^{jk}_w p^j_w p^k_w \nonumber\\
	& &
	+ (k^{{\rm NR}}_{\sigma\phi})_w^j \sigma^j_w +(k^{{\rm NR}}_{\sigma\phi pp})_w^{jkl} \sigma^j_w p^k_w p^l_w \Big] \frac{\vec{g}}{m_{atom}},
	\label{operator}
\end{eqnarray}
where $w$ stands for e, n, and p. The coefficients $(k^{{\rm NR}}_{\phi})_w$ and $(k^{{\rm NR}}_{\phi pp})^{jk}_w$ denote the spin-independent couplings to gravity via the gravitational potential $\phi$. The coefficients $(k^{{\rm NR}}_{\sigma\phi})_w^j$ and $(k^{{\rm NR}}_{\sigma\phi pp})_w^{jkl}$ denote the spin-dependent couplings to gravity via $\phi$.
For typical atoms, the mean momentum squared is of order
$\langle\vec{p}^{\, 2}\rangle_e \simeq10^{-11}$ GeV$^2$ and $\langle\vec{p}^{\, 2}\rangle_p 
\simeq \langle\vec{p}^{\, 2}\rangle_n\simeq10^{-2}$ GeV$^2$. So when calculating the terms quadratic in momenta, contributions from electrons are neglected.

\subsection{Calculation for the $^{85}$Rb and $^{87}$Rb atoms}

For our purpose, we will discuss how to calculate Eq. (\ref{operator}) for the 5S$_{1/2}$ $|$F=2, $F_z$=0$\rangle$ and 5S$_{1/2}$ $|$F=3, $F_z$=0$\rangle$ of $^{85}$Rb atoms, and for the 5S$_{1/2}$ $|$F=1, $F_z$=0$\rangle$ and 5S$_{1/2}$ $|$F=2, $F_z$=0$\rangle$ of $^{87}$Rb atoms. Note that $F$ denotes the atomic spin number, and $F_z$ denotes the z-component of the atomic spin. 

For both $^{85}$Rb and $^{87}$Rb atoms, there is a single valence electron in the $5^2S_{1/2}$ level,
with the total electronic angular momentum $J^e=1/2$. The nucleus of $^{85}$Rb atoms has nuclear spin $I=5/2$ with orbital angular momenta $L^p=3$, so $I=L^p-1/2$ \cite{PhysRevC.10.2117}. Then, the atomic spin of $^{85}$Rb is given by $F=I \pm J^e=3$ or 2.
The nucleus of $^{87}$Rb atoms has nuclear spin $I=3/2$ with orbital angular momenta $L^p=1$, so $I=L^p+1/2$ \cite{PhysRevC.12.1516}. So, the atomic spin of $^{87}$Rb is given by $F=I \pm J^e=2$ or 1.
In the Schmidt model \cite{Blatt1979}, the nuclear properties for both $^{85}$Rb and $^{87}$Rb atoms are assigned to a single valence proton because the number of neutrons (48 and 50, respectively) is very close to a magic number. 

One can express the atomic state as the tensor product of the valence electron state and  the valence proton state, 
\begin{equation*}
	|\mathcal{R}, F, F_z\rangle = \langle F, F_z | J^e, J^e_z, I, I_z \rangle
	|\mathcal{R}', J^e, J^e_z\rangle |\mathcal{R}'', I, m_I\rangle,
\end{equation*}
where $\langle F, F_z | J^e, J^e_z, I, I_z \rangle$ is a Clebsch-Gordan coefficient. $\mathcal{R}$, $\mathcal{R}'$, and $\mathcal{R}''$ denote the radial dependence. Then, it is straightforward to write down
\begin{widetext}
	\begin{eqnarray*}
		^{85}{\rm Rb}|3, 0\rangle &=& \sqrt{\frac{1}{2}}\left(-\sqrt{\frac{4}{7}}Y_{3,-1} |{1 \over 2}\rangle^p +\sqrt{\frac{3}{7}}Y_{3,0} |{-1 \over 2}\rangle^p\right)|{1 \over 2}\rangle^e
		+\sqrt{\frac{1}{2}}\left(-\sqrt{\frac{3}{7}}Y_{3,0} |{1 \over 2}\rangle^p +\sqrt{\frac{4}{7}}Y_{3,1} |{-1 \over 2}\rangle^p\right)|{-1 \over 2}\rangle^e\\
		^{85}{\rm Rb}|2, 0\rangle &=& -\sqrt{\frac{1}{2}}\left(-\sqrt{\frac{4}{7}}Y_{3,-1} |{1 \over 2}\rangle^p +\sqrt{\frac{3}{7}}Y_{3,0} |{-1 \over 2}\rangle^p\right)|{1 \over 2}\rangle^e
		+\sqrt{\frac{1}{2}}\left(-\sqrt{\frac{3}{7}}Y_{3,0} |{1 \over 2}\rangle^p +\sqrt{\frac{4}{7}}Y_{3,1} |{-1 \over 2}\rangle^p\right)|{-1 \over 2}\rangle^e\\
		^{87}{\rm Rb}|2, 0\rangle &=& \sqrt{\frac{1}{2}}\left(\sqrt{\frac{1}{3}}Y_{1,-1} |{1 \over 2}\rangle^p +\sqrt{\frac{2}{3}}Y_{1,0} |{-1 \over 2}\rangle^p\right)|{1 \over 2}\rangle^e
		+\sqrt{\frac{1}{2}}\left(\sqrt{\frac{2}{3}}Y_{1,0} |{1 \over 2}\rangle^p +\sqrt{\frac{1}{3}}Y_{1,1} |{-1 \over 2}\rangle^p\right)|{-1 \over 2}\rangle^e\\
		^{87}{\rm Rb}|1, 0\rangle &=& -\sqrt{\frac{1}{2}}\left(\sqrt{\frac{1}{3}}Y_{1,-1} |{1 \over 2}\rangle^p +\sqrt{\frac{2}{3}}Y_{1,0} |{-1 \over 2}\rangle^p\right)|{1 \over 2}\rangle^e
		+\sqrt{\frac{1}{2}}\left(\sqrt{\frac{2}{3}}Y_{1,0} |{1 \over 2}\rangle^p +\sqrt{\frac{1}{3}}Y_{1,1} |{-1 \over 2}\rangle^p\right)|{-1 \over 2}\rangle^e
	\end{eqnarray*}	
\end{widetext}
where $Y_{l, m}$'s are the spherical harmonics. The superscripts $e$ and $p$ stand for the electron and proton, respectively. After some calculation, it is easy to find out the expectation values of Eq. (\ref{operator}) for the four states in the laboratory frame, which are
\begin{widetext}
	\begin{eqnarray}
		\left. {\delta g \over g}\right|_{{\rm ^{85}Rb}|3\rangle}= \left. {\delta g \over g}\right|_{{\rm ^{85}Rb}|2\rangle}&=&{1\over m_{85}}\left[ (k^{{\rm NR}}_{\phi})_e+(k^{{\rm NR}}_{\phi})_p+\left( {9\over 35}\left((k^{{\rm NR}}_{\phi pp})_p^{xx}+(k^{{\rm NR}}_{\phi pp})_p^{yy}\right)+{17\over 35}(k^{{\rm NR}}_{\phi pp})_p^{zz}\right) \left\langle \vec{p}^{\,2} \right\rangle_p \right] \\
		\label{sme85}
		\left.{\delta g \over g}\right|_{{\rm ^{87}Rb}|2\rangle} =\left. {\delta g \over g}\right|_{{\rm ^{87}Rb}|1\rangle}&=&{1\over m_{87}}\left[ (k^{{\rm NR}}_{\phi})_e+(k^{{\rm NR}}_{\phi})_p+\left( {4\over 15}\left((k^{{\rm NR}}_{\phi pp})_p^{xx}+(k^{{\rm NR}}_{\phi pp})_p^{yy}\right)+{7\over 15}(k^{{\rm NR}}_{\phi pp})_p^{zz}\right) \left\langle \vec{p}^{\,2} \right\rangle_p \right]
		\label{sme87}
	\end{eqnarray}	
\end{widetext}
where m$_{85}$=79.1 GeV, and m$_{87}$=80.9 GeV are the atomic masses of $^{85}$Rb and $^{87}$Rb atoms \cite{PhysRevA.82.042513}.

\bibliography{wepspin2024.bib}

\end{document}